\documentclass[twocolumn,showpacs,nofootinbib]{revtex4-1}

\usepackage{amsmath}
\usepackage{graphicx}
\usepackage{bm}
\usepackage{pbox}
\usepackage{float}
\usepackage[utf8]{inputenc}
\usepackage[T1]{fontenc}

\def\to{\rightarrow}

\def\al{\alpha} \def\be{\beta}  
   
   \def\ka{\kappa}

 \def\Om{\Omega}  \def\cl{{\cal L}}
\def\fr#1#2{{{#1} \over {#2}}}

\def\half{{\textstyle{1\over 2}}} \def\frac#1#2{{\textstyle{{#1}\over
			{#2}}}} 
\def\lsim{\mathrel{\rlap{\lower4pt\hbox{\hskip1pt$\sim$}}
		\raise1pt\hbox{$<$}}}
\def\gsim{\mathrel{\rlap{\lower4pt\hbox{\hskip1pt$\sim$}}
		\raise1pt\hbox{$>$}}} \def\sqr#1#2{{\vcenter{\vbox{\hrule
				height.#2pt \hbox{\vrule width.#2pt height#1pt \kern#1pt \vrule
					width.#2pt} \hrule height.#2pt}}}}

\def\beq{\begin{equation}} \def\eeq{\end{equation}}
\def\beqa{\begin{eqnarray}} \def\eeqa{\end{eqnarray}}
\def\beal{\begin{align}} \def\eal{\end{align}}
\def\ca{\mathcal{A}} \def\cb{\mathcal{B}}  \def\cd{\mathcal{D}}    \def\cl{\mathcal{L}} 

\usepackage{hyperref}

\begin{document}

\title{Cosmological dynamics of a class of non-minimally coupled models of gravity}

\author{R.P.L. Azevedo}
\email[]{up201109025@fc.up.pt}
\author{J. P\'aramos}
\email[]{jorge.paramos@fc.up.pt}
\affiliation{Departamento de F\'\i sica e Astronomia and Centro de F\'isica do Porto, \\Faculdade de Ci\^encias da Universidade do Porto,\\Rua do Campo Alegre 687, 4169-007 Porto, Portugal}

\date{\today}

\begin{abstract}

In this work a new non-minimally coupled model is presented, where a generic function $f(R)$ of the scalar curvature factors the usual Einstein-Hilbert action functional, motivated by relevant results obtained from similar models. Its cosmological dynamics are derived and the possibility of attaining a phase of accelerated expansion is assessed. To further probe the possible implications of the model, a dynamical system formulation is established, and used to assess the scenarios where $f(R)$ assumes a power-law or exponential form.
\end{abstract}


\maketitle

\section{Introduction}\label{sec:intro}

Albert Einstein's General Relativity (GR) has served as the framework for the development of the so called standard model of cosmology. It is the simplest theory that relates matter and the curvature of spacetime, and by far the one with most experimental support \cite{experimental1,experimental2}, from the prediction of the precession of Mercury's perihelion to the recent detection of gravitational wave production by black hole binaries \cite{ligo}.

Despite this backing, when coupled only with baryonic matter, GR still fails to account for more recent observations of the Universe. Comparisons of the rotational speed and mass of galaxies  as predicted by GR and as measured via electromagnetic radiation do not appear to match, as if there were some missing mass from our calculations. Moreover, in the past two decades observations of supernovae have signalled that the Universe is expanding at an accelerating rate \cite{dark}. To address these flaws, the $\Lambda$CDM model was formulated, consisting of a universe evolving under GR and with the addition of dark energy, represented by a cosmological constant $\Lambda$ with negative Equation of State (EOS) and that is responsible for this accelerated expansion, and Cold Dark Matter (CDM), a non-baryonic type of matter that either does not interact electromagnetically or has a vanishingly small interaction, which is responsible for this missing mass. This model is also supplemented by an inflationary scenario based on a scalar field to explain the early exponential expansion of the Universe.

An alternative to this solution is to assume that GR is incomplete, prompting other models to appear and attempt to explain this large scale behaviour. Among the most prominent are the so-called $f(R)$ theories \cite{felice,fR1,fR2,fR3,fR4,fR5,fR6}, where the Einstein-Hilbert action is replaced by a nonlinear function of the scalar curvature, and models that present non-minimal couplings (NMC) between matter and curvature \cite{early1,early2,early3,early4,Bertolami2007gv}. Some of these models have been shown to be able to mimic dark matter \cite{NMCDM1,NMCDM2,NMCDM3,NMCDM4} or dark energy \cite{Bertolami2010cw,Bertolami2011fz,Bertolami2013uwl}, and explain post-inflationary preheating \cite{Bertolami2010ke} and cosmological structure formation \cite{NMCS1,NMCS2,NMCS3}.

Previous attempts at solving these cosmological problems using a NMC model have resorted to a coupling between curvature and a scalar field \cite{singleNMCDE1,singleNMCDE2,singleNMCDE3,repetido,singleNMCCI1,singleNMCCI2, singleNMCHI1,singleNMCHI2,singleNMCHI3}, but did not extend this coupling to the baryonic matter content. More recently, a dynamical system analysis approach was used to analyse a model that incorporated both $f(R)$ theories and a NMC with the baryonic matter content \cite{rafael2014}.

Taking this research background into account, in this work we do a dynamical system approach on a particular group of NMC theories, represented by the Lagrangian density $f(R)(\ka R+\cl)$, which presents an both an appealing form and interesting behaviour. This method, on which we will elaborate further in the following sections, allows us to check for the existence of solutions to the cosmological equations, and to analyse their stability. Other similar studies, albeit in a different context, can be found in Refs. \cite{Azizi2014qsa,Carloni2004kp,Carloni2007br,fttg,Azevedo:2016}.

This work is organized as follow: the model under scrutiny is discussed in Sec. \ref{sec:model}; the derivation of the corresponding dynamical system is found in Sec. \ref{sec:system}; the results and respective discussion of an exponential and a power law models can be found in Secs. \ref{sec:detfRexp} and \ref{sec:detfRpower}, respectively. Finally, conclusions are drawn in Sec. \ref{sec:conclusion}.

\section{The Model}\label{sec:model}

We consider a NMC theory that follows from $f(R)$ theories with a coupling between a generic function of the scalar curvature $R$ and the standard Einstein-Hilbert Lagrangian, embodied in the action
\beq\label{actiondetfR}
S=\int d^4x \sqrt{-g}f(R)(\kappa R +\cl),
\eeq
where $\cl$ is the matter Lagrangian density, $g$ is the determinant of the metric and $\kappa=c^4/(16\pi G)$. GR is recovered by setting $f(R) = 1$.

This type of coupling could be looked upon as a more elegant extension of the standard minimal coupling between matter and curvature \cite{Bertolami2007gv}: instead of modifying the coupling term $\sqrt{-g}\cl$ in particular, one generally chooses to modify $\sqrt{-g}$ and maintain the Einstein-Hilbert term $\kappa R +\cl$. In this sense, the modification above can be viewed as a geometrically inspired extension of GR, where the measure $\sqrt{-g}$ is generalized to also depend on the scalar curvature.

In Ref. \cite{Castel:2014}, the authors adopt a model embodying both a non-linear function of the scalar curvature and a NMC between curvature and matter, given by the action
\beq
S=\int d^4x \sqrt{-g}[f_1(R) +f_2(R)\cl],
\eeq
\noindent with
\beqa\label{f1f2taylor}
f_1(R)&=&\kappa\left(R+{R^2\over6m^2}\right)+\mathcal{O}(R^3),\\ \nonumber f_2(R)&=& 1 + {R\over 6M^2}+\mathcal{O}(R^2),
\eeqa
where $m$ and $M$ are characteristic mass scales. As is shown in the cited paper, using an adequate metric that describes the spacetime around a spherical star like the Sun, one is able to identify a Newtonian potential with an additional Yukawa term,
\beq\label{yukawa}
U(r)=-{GM_S\over r}\left[1+\alpha A(m,R_S)e^{-r/\lambda}\right],
\eeq
where $M_S$ and $R_S$ are respectively the mass and radius of the star, $A(m,R_S)$ is a form factor, $\lambda=1/m$ and $\alpha \propto 1-(m/M)^2$.

Solar system tests in this framework suggest that $|\alpha| \ll 1$ for $\lambda $ ranging from submillimiter scales up to $100$ AU \cite{adelberger}: thus, in pure $f(R)$ models (obtained by setting $M \to \infty$), this additional Yukawa contribution has the same strength as gravity, $\alpha = 1$, and accordingly $\lambda$ must lie outside the cited range --- although, if one assumes a non-vanishing density outside the central body, a chameleon effect may arise where the dynamical impact of a non-linear $f(R)$ can be hidden from local tests of gravity due to the reduction of its Compton wavelength in regions of deep gravitational potential wells \cite{HuSawicki,Brax}. 

Conversely, the presence of a NMC may avoid clashing with experimental constraints if both mass scales are very close, $M \sim m$; if one simply assumes that both functions $f_i(R)$ share the same mass scale, $M = m$, then one has $f_1(R) = \ka f_2(R) R$ and the action may be written as 
\beq
S=\int d^4x \sqrt{-g} \left( 1 + {R\over 6M^2} \right) (\kappa R +\cl).
\eeq
Thus, one concludes that the above factorisation of the Einstein-Hilbert action leads to vanishing first-order effects (see Ref. \cite{March:2016xav} for a second-order treatment of the remaining dynamics).

Furthermore, in a cosmological context \cite{rafael2014}, a factorisation of the form of action (\ref{actiondetfR}) with $f(R) \sim 1 + (R/M^2)^n$ is shown to allow for a matter-dominated universe that behaves as if GR was valid, effectively concealing the effect of the additional contribution arising from a non-trivial $f(R)$ function (although other cosmological fixed points arise which showcase the additional dynamics).

The above cases illustrate the interesting consequences of assuming the action (\ref{actiondetfR}), which will be explored in the following sections.

\subsection{Cosmological Dynamics}

A null variation of the action \eqref{actiondetfR} with respect to the metric gives us the field equations
\beq\label{field_detfR}
FG_{\mu\nu}={1\over2}f T_{\mu\nu} + \Delta_{\mu\nu}F+ \half g_{\mu\nu}\kappa f -{1 \over 2}g_{\mu\nu} RF,
\eeq
where $F=\kappa f(R)+f'(R)(\kappa R+\cl )$ and primes denote differentiation with respect to the scalar curvature.

where $G_{\mu\nu}\equiv R_{\mu\nu}-g_{\mu\nu}R/2$ is the Einstein tensor and $T_{\mu\nu}$ is the matter energy-momentum tensor, defined as
\begin{equation}\label{energy-mom}
T_{\mu\nu} = - \fr{2}{\sqrt{-g}}\fr{\delta\left(\sqrt{-g}\mathcal{L}\right)}{\delta g^{\mu\nu}}.
\end{equation}

The Bianchi identities imply the noncovariant conservation law \cite{Bertolami2007gv}
\begin{equation}\label{conservNMC}
\nabla^\mu T_{\mu\nu} = {f'\over f}\left(g_{\mu\nu}\cl-T_{\mu\nu}\right)\nabla^\mu R.
\end{equation}

Considering the Cosmological Principle, {\it i.e.} that the Universe is homogeneous and isotropic, and assuming spatial flatness, it can be well described via a Friedmann-Lema\^itre-Robertson-Walker (FLRW) metric, represented by the line element
\begin{equation}\label{line}
ds^2=-dt^2+a^2(t)dV^2,
\end{equation}
\noindent where $a(t)$ is the scale factor and $dV$ is the volume element in comoving coordinates. From a cosmological standpoint, the matter content of the Universe can be described as a perfect fluid, with energy-momentum tensor
\begin{equation}\label{fluid}
T^{\mu\nu}=(\rho+p)u^\mu u^\nu + p g^{\mu\nu},
\end{equation}	
\noindent derived from the Lagrangian density $\mathcal{L}=-\rho$ (see Refs.~\cite{fluid1,fluid2,fluid3} for a discussion), where $\rho$ and $p$ are, respectively, the energy density and pressure of the perfect fluid, and $u^\mu$ is its four-velocity, with the normalization condition $u_\mu u^\mu=-1$. The pressure and energy density are considered to obey a barotropic equation of state (EOS) $p=w\rho$, where $w$ is the EOS parameter; since this work is focused on alternative explanations for dark energy, we exclude $w\neq -1$.

Note that, even though Eq.~\eqref{conservNMC} implies that energy is not generally conserved, the used metric \eqref{line} and energy-momentum tensor \eqref{fluid} make the right side of the conservation equation vanish, and one obtains the usual continuity equation
\begin{equation}\label{contin}
\dot{\rho}+3H(1+w)\rho=0,
\end{equation}	
\noindent where $H\equiv\dot{a}/a$ is the Hubble parameter.

Introducing the metric \eqref{line} and energy-momentum tensor \eqref{fluid} into the field Eqs. \eqref{field_detfR} one  obtains the modified Friedmann and Raychaudhuri equations,  respectively
\beq\label{fried_detfR}
H^2={1\over3F}\bigg[{1\over2}R(F-\kappa f)-3HF'\dot{R}+{1\over2}f\rho -9H^2(1+w)f'\rho\bigg],
\eeq
\beq \label{ray_detfR}
2\dot{H}+3H^2={1\over2F}\left[R(F-\kappa f)-2\ddot{F}-4H\dot{F}
-f w\rho\right],
\eeq
with $F'\equiv 2\kappa f' + f''(\kappa R-\rho)$.

\subsection{De Sitter Solution}\label{deSitter}

An interesting exercise is to determine under which conditions these equations result in a de Sitter universe, {\it i.e.} an exponential scale factor $a(t)=e^{H_0 t}$. Eq. \eqref{Ricci} leads to a constant Ricci scalar $R_0=12 H_0^2\neq 0$, and, and shown below, the modified Friedmann \eqref{fried_detfR} and Raychaudhuri \eqref{ray_detfR} equations then posit two scenarios, depending on the value of the energy density $\rho$.

\subsubsection{Solutions with an empty universe}\label{emptydeSitter}

We define $f_0 \equiv f(R_0)$ and $f'_0 \equiv f'(R_0)$; assuming that the universe is devoid of any kind of matter, $\rho=0$, one has $ F = \kappa (f_0 + f_0' R_0)$, so that Eqs. \eqref{fried_detfR} and \eqref{ray_detfR} both read
\beq f_0 = f'_0 R_0, \eeq
\noindent thus yielding a condition for the allowed form of $f(R)$.

If the latter has {\it e.g.} an exponential form, this yields
\beq \label{rho0fR}
f_0=\exp\left({R_0\over M^2}\right)\rightarrow R_0=M^2\rightarrow H_0={M\over 2\sqrt{3}},
\eeq
\noindent a result verified in Section \ref{sec:detfRexp}.

If a power-law behaviour is assumed instead, $f(R) \sim R^n$, the above condition requires that $n=1$, {\it i.e.} a linear form --- as shall be shown in Section \ref{sec:detfRpower}.

\subsubsection{Solutions with a non-empty Universe}\label{nonemptydeSitter}

On the other hand, if $\rho \neq 0$, one has
\beqa \dot{F} &=& -f_0' \dot{\rho} = 3f_0' H (1+w)\rho \to \\ \nonumber \ddot{F} &=& - 9 f_0'H_0^2 (1+w)^2\rho, \eeqa
\noindent having used the conservation Eq. (\ref{contin}), so that Eqs. \eqref{fried_detfR} and \eqref{ray_detfR} read
\beqa
\ka( f_0 - f_0'R_0 ) R_0 &=& [2f_0 - (4+3w)f_0' R_0]\rho \\ \nonumber &=& - w [2f_0 - (4+3w)f_0' R_0 ] \rho.
\eeqa
As a sanity check notice that, in the case of GR, $ f(R) = 1 $ naturally implies a fluid behaving as a Cosmological Constant, $w = -1$ and $\rho = \rho_{\Lambda} \equiv \ka R_0 / 2$.
For non-trivial forms of $f(R)$, and since the scalar curvature is constant while the energy density of the assumed baryonic matter decreases, the above implies that both sides of the relation should vanish: this can only be attained if
\beq f_0 = f'_0 = 0, \eeq
\noindent leading to the conclusion that a regime of De Sitter expansion with a non-negligible matter contribution requires a much more stringent condition than if the energy density vanishes.

In particular, neither the exponential nor the power-law forms for $f(R)$ assumed in the preceding paragraphs can fulfil this criteria, and no de Sitter solutions with matter are attainable (as shown in Section \ref{sec:detfRexp} and \ref{sec:detfRpower}).

\section{Dynamical System}\label{sec:system}

One can study solutions to the field equations by analysing the dynamical system that results from the modified Friedmann and Raychaudhuri equations, written in the terms of the dimensionless variables
\begin{align}\label{variables}
&x=-{F' \dot{R}\over F H}~~~~,~~~~y={R \over 6H^2}~~~~,~~~~z=-{\kappa fR\over 6F H^2}, \nonumber \\
&\Omega_1={f \rho \over 6 F H^2}~~~~,~~~~\Omega_2=-{3(1+w)f' \rho \over F},
\end{align}
\noindent such that the modified Friedmann equation can be read
\beq\label{mfe}
1=x+y+z+\Omega_1+\Omega_2,
\eeq
\noindent acting as a restriction on the phase space.
The quantities $\dot{F}/\left(F H\right)$ and $\ddot{F} /\left(F H^2\right)$ are useful in the subsequent derivations, so one should write them as functions of the variables \eqref{variables}:
\begin{align}\label{fdot}
&{\dot{F} \over F H} =-(x+\Omega_2),\\ \nonumber 
&{\ddot{F} \over F H^2} = (x+\Omega_2)(x+\Omega_2+2-y)-{dx \over dN}- {d\Omega_2 \over dN},
\end{align}
where $N$ is the number of $e$-folds.

Rewriting the modified Raychaudhuri Eq. \eqref{ray_detfR} as a function of the dimensionless variables defined above, the following relation may be obtained
\begin{equation}\label{dimensionlessRaychaudhuri}
{dx \over dN} + {d\Omega_2 \over dN} = (x+\Omega_2)(x+\Omega_2-y)-y-3z+3w\Omega_1-1.
\end{equation}
\noindent We can write this relation more explicitly by differentiating $\Omega_2$ and using the continuity equation to obtain
\begin{equation}\label{evoomega2}
{d\Omega_2\over dN} = \Omega_2\left[x\left(1-{\gamma \over \alpha}\right)- 3(1+w)+\Omega_2\right],
\end{equation}
\noindent and we obtain the first equation of our dynamical system, equivalent to the Raychaudhuri equation,
\begin{align}\label{evox}
{dx \over dN} =& x\left[x-y+\Omega_2\left(1+{\gamma \over \alpha}\right)\right] -1 -y -3z +3w\Omega_1 \nonumber \\
& +  \Omega_2\left[3(1+w)-y\right],
\end{align}
where we have made use of the dimensionless parameters
\beq\label{dimpar_detfR}
\alpha(R,\rho)={F'R\over F},~~~~ \beta(R)={f'R\over f}+1,~~~~ \gamma(R)={f''R\over f'}.
\eeq

Following from the conservation law \eqref{contin}, one can derive the variables \eqref{variables} with respect to the number of $e$-folds and obtain the autonomous system
\begin{equation}\label{system}
\begin{cases}
{dx \over dN} = &x\left[x-y+\Omega_2\left(1+{\gamma \over \alpha}\right)\right] -1 -y -3z  \\
&+ 3w\Omega_1 +\Omega_2\left[3(1+w)-y\right] \\
{dy \over dN} = &y\left[2(2-y)-{x \over \alpha}\right] \\
{dz \over dN} = &z\left[x\left(1-{\beta \over \alpha} \right)+\Omega_2+2(2-y)\right] \\
{d\Omega_1 \over dN} = &{\Omega_2 xy \over 3\alpha(1+w)} +\Omega_1\left(1-3w+x+\Omega_2-2y\right) \\
{d\Omega_2 \over dN} = &\Omega_2\left[x\left(1-{\gamma \over \alpha}\right)- 3(1+w)+\Omega_2\right]
\end{cases}.
\end{equation}

Since the Raychaudhuri equation can be calculated by differentiating the Friedmann equation, and it is also equivalent to the relation for $dx/dN$, the relation
\beq\label{extra_cond}
{dx\over dN}+{dy\over dN}+{dz\over dN}+{d\Omega_1\over dN}+{d\Omega_2\over dN}=0
\eeq
\noindent must hold. Fortunately, instead of vanishing trivially, Eq.~\eqref{extra_cond} yields
\beq\label{extra_cond2}
y\left[{\Omega_2\over3(1+w)}-1\right]=z\beta.
\eeq
Relation \eqref{extra_cond2} and the Friedmann equation \eqref{mfe} act on the system \eqref{system} as algebraic constraints, and allow us to reduce its dimensionality. Eliminating $\Omega_1$ and $\Omega_2$, we are left with
\begin{align}\label{system_redux}
\begin{cases}
{dx \over dN} &= x\left[x-y+3(1+w)\left(1+{z\over y}\beta\right)\left(1+{\gamma \over \alpha}\right)-3w\right]+\\ 
&2(2+3w)(2-y) -3(1+w)z \left[ 1 + \left(1 - {3 \over y}\right)\beta \right] \\
{dy \over dN} &= y\left[2(2-y)-{x \over \alpha}\right] \\
{dz \over dN} &= z\Big[x\left(1-{\beta \over \alpha} \right)+3(1+w)\left(1+{z\over y}\beta\right)+2(2-y)\Big] \\
\end{cases}.
\end{align}

Solving this system generally also requires writing the scalar curvature $R$ and the energy density $\rho$ as functions of the variables, which can be done recurring to the definition of the variables themselves. Specifically, one finds the scalar curvature by inverting
\beq\label{reldet1}
{f'(R)R\over f(R)}=-{\Omega_2 y\over 3(1+w)\Omega_1}=-{y+z\beta(R) \over \Omega_1},
\eeq
and the energy density from
\beq\label{reldet2}
\rho(y,z,\Omega_1)=-\kappa R(y,z,\Omega_1){\Omega_1\over z}.
\eeq

\subsection{De Sitter Solution}

Following subsection \ref{deSitter} and the relations given in the Appendix, one may now impose on Eqs. (\ref{system_redux}) the condition $y=2$ and $x \sim \dot{R} = 0 $, corresponding to a de Sitter phase of exponential evolution of the scale factor, so that the scalar curvature and Hubble parameter are constant and related by $ R= R_0 = 12H_0^2$.

The relation for $dy/dN$ is then trivially satisfied, while the relations for $dx/dN$ and $dz/dN$ read
\beq \label{dxdNdeSitter} \begin{cases} 0 &= z (\beta-2) \\ 0&= z(2+z \be ) \end{cases} \to z = \{-1, 0 \}. \eeq

Thus, two possibilites arise:

\begin{itemize}
\item Case $z=-1$:

Eq. (\ref{dxdNdeSitter}) implies that we must have $\beta = 2$: from its definition (\ref{dimpar_detfR}), this reads
\beq \label{betacrit} \beta = {f'_0R_0 \over f_0} + 1 = 2 \to f'_0R_0 = f_0,\eeq
the condition derived in paragraph \ref{emptydeSitter} for a de Sitter expansion in a waterless scenario.

Notice that the algebraic relation (\ref{extra_cond2}) yields $\Om_2 = 0$ and the Friedmann constraint (\ref{mfe}) further implies that $\Om_1 = 0 $, reflecting a vanishing energy density.

\item Case $z=0$:

From definition (\ref{variables}), we see that $z=0$ requires $f_0 = 0 $. Also, the algebraic relation (\ref{extra_cond2}) leads to $\Om_2 = 3(1+w)$; using the Friedmann constraint (\ref{mfe}) finally yields $\Om_1 = -(4+3w)$.

Since $f_0 = 0$, the definitions (\ref{variables}) thus imply that $f_0=f'_0=0$ --- as obtained in paragraph \ref{nonemptydeSitter} for a de Sitter expansion in a universe with a non-vanishing matter contribution.

\end{itemize}

The above serves not only to corroborate the previous findings of Subsection \ref{deSitter} but, more crucially, to attest that the choice of variables embodied in Eq. (\ref{variables}) does not miss out the relevant dynamics (see, however, the discussion in the Conclusions).
\section{Exponential $f(R)$}\label{sec:detfRexp}

We now proceed to study a model with 
\beq\label{dfRexp}
f(R)=\exp\left({R\over M^2}\right),
\eeq
where $M$ is a characteristic mass scale; this theory collapses to GR for large $M$ or small $R$. The exponential form of the theory makes it very straightforward to calculate the dimensionless parameters \eqref{dimpar_detfR},
\beq\label{parexpfR}
\al= {2R \over \kappa R-\rho}+{R\over RM^2}~~,~~\beta=1+{R\over M^2}~~,~~\gamma={R\over M^2}.
\eeq
Due to the complexity of the fixed point solutions, we constrained the results to only include dust, {\it i.e.} pressureless matter with $w=0$. The fixed points associated with this function can be found in Table \ref{fix_expfR}. It should be noted that the values of the scalar curvature and energy density naturally depend on the mass scale $M^2$.
\begin{table*}\centering
	\caption{Fixed points and corresponding solutions for an exponential $f(R)$.\label{fix_expfR}}
	\begin{tabular*}{\textwidth}{@{\extracolsep{\fill}}lcccccc}
		\hline \hline
		Point&$(x,y,z,\Omega_1,\Omega_2)$	&$(R/M^2,\rho / \kappa M^2)$	&$a(t)$	&$\rho(t)$	&$q$\\ \hline 
		$\mathcal{A}$	&$\left(0,2,-1,0,0\right)$	&$\left(1,0 \right)$	&$e^{H_0 t}$ 	&$\rho_0 e^{-3H_0 t}$	&$-1$\\
		
		$\mathcal{B}$	&$\left(0.552,1.605,-1.244,0.327,-0.241\right)$	&$\left(0.394,0.104\right)$	&$\left({t\over t_0}\right)^{2.533}$ 	&$\rho_0 \left({t\over t_0}\right)^{-7.598} $	&$-0.605$\\
			
		$\mathcal{C}$	&$\left(0.601,1.266,-1.140,0.558,-0.285\right)$	&$\left(0.216,0.105\right)$	&$\left({t\over t_0}\right)^{1.363}$ 	&$\rho_0 \left({t\over t_0}\right)^{-4.089}$	&$-0.266$\\
		\hline \hline
	\end{tabular*}
\end{table*}

\subsection*{Point $\mathcal{A}$}\label{subsec:expfRA}
The first point corresponds to a stable de Sitter universe with vanishing energy density at $t\rightarrow\infty$, whose expansion rate can be calculated from the definition
\beq\label{expratefRA}
z=-{\kappa fR \over 6FH^2}=-1\rightarrow H_0={M\over 2\sqrt{3}}.
\eeq

It should be noted that this point is always attained, and stable, for $w\geq 0$, so that an exponential form of the coupling is capable of generating such solutions, regardless of the type of barotropic matter considered.
\subsection*{Points $\mathcal{B}$ and $\mathcal{C}$}\label{subsec:expfRB}
Both points are stable and present negative deceleration parameters, and as such are both good candidates for dark energy (particularly point $\mathcal{B}$, since $q = -0.605$ is quite close to the present value). They differ from point $\mathcal{A}$ in that they do not require a vanishing energy density.


\section{Power Law $f(R)$}\label{sec:detfRpower}

We now consider a power law model
\beq\label{dfRpower}
f(R)=\left({R\over M^2}\right)^n,
\eeq
where $M$ is mass scale, and that approaches GR if $M$ is large or $n$ is very small. In this case the parameters \eqref{dimpar_detfR} take the form
\beq\label{parpowerfR0}
\al=n \left[ 1 + { \rho \over (n+1)\ka R -n\rho }\right]~~,~~\beta=n+1~~,~~\gamma=n-1.
\eeq
The fixed points and solutions associated with this function can be found in Table \ref{fix_powerfR0}.

\begin{table*}\centering
	\caption{Fixed points and solutions for a power law $f(R)$.\label{fix_powerfR0}}
	\begin{tabular*}{\textwidth}{@{\extracolsep{\fill}}lccc}
		\hline \hline
		Point&$(x,y,z,\Omega_1,\Omega_2)$	&$a(t)$	&$q$\\ \hline 
		$\mathcal{A}$	&\pbox{20cm}{
			$\Bigg(
			3 n (1+w)\left( 1 + n { 6 n (1+w)+3 w+5 \over 3 w-1}\right),
			{1 \over 2} (1-3 w),
			n \left[3 (1+w)n-{1 + 3 w\over 2}\right]+{3 w-1 \over 2},$\\
			$1-n \left[3 (1+w)n+{3 w+5\over 2}\right],
			-3 (1+w)n { n [6 n (1+w)+3 w+5]-2 \over 3 w-1}
			\Bigg)$}
		&$	\left({t \over t_0}\right)^{2\over 3(1+w)}$ 	
		&$ {1 \over 2} (1+3w)$\\
		
		$\mathcal{B}~~(n\neq 1)$	&$\left(
		-1+\fr{3}{2 n+1},
		2-\fr{1}{n}+\fr{3}{2 n+1},
		\fr{1}{n}-\fr{6}{2 n+1},
		0,
		0
		\right)$ 
		&$\left({t \over t_0}\right)^{n(2n+1)\over 1-n} $ 	
		&$-1+\fr{1}{n}-\fr{3}{2 n+1}$\\
		
		$\mathcal{B}~~(n=1)$	&$\left(
		0,
		2,
		-1,
		0,
		0
		\right)$ 
		&$e^{H_0 t}$ 	
		&$-1$\\
		
		$\mathcal{C}$	&$\left(
		0,
		2,
		0,
		-(4+3w),
		3(1+w)
		\right)$
		&${\rm const.}$ 	
		&\\
		
		$\mathcal{D}$	&$\left(
		\fr{6 n w}{1-2 n}-4,
		n\fr{4 n+3 w-2}{(n-1) (2 n-1)},
		0,		
		-\fr{4 n+3 w-2}{(n-1) (2 n-1)},
		3 (1+w)		
		\right)$
		&${\rm const.}$ 	
		&	\\
		\hline \hline
	\end{tabular*}
\end{table*}
\subsubsection*{Point $\mathcal{A}$}\label{subsubsec:powerfRA}
\begin{figure}
	\centering
	\includegraphics[width=\columnwidth]{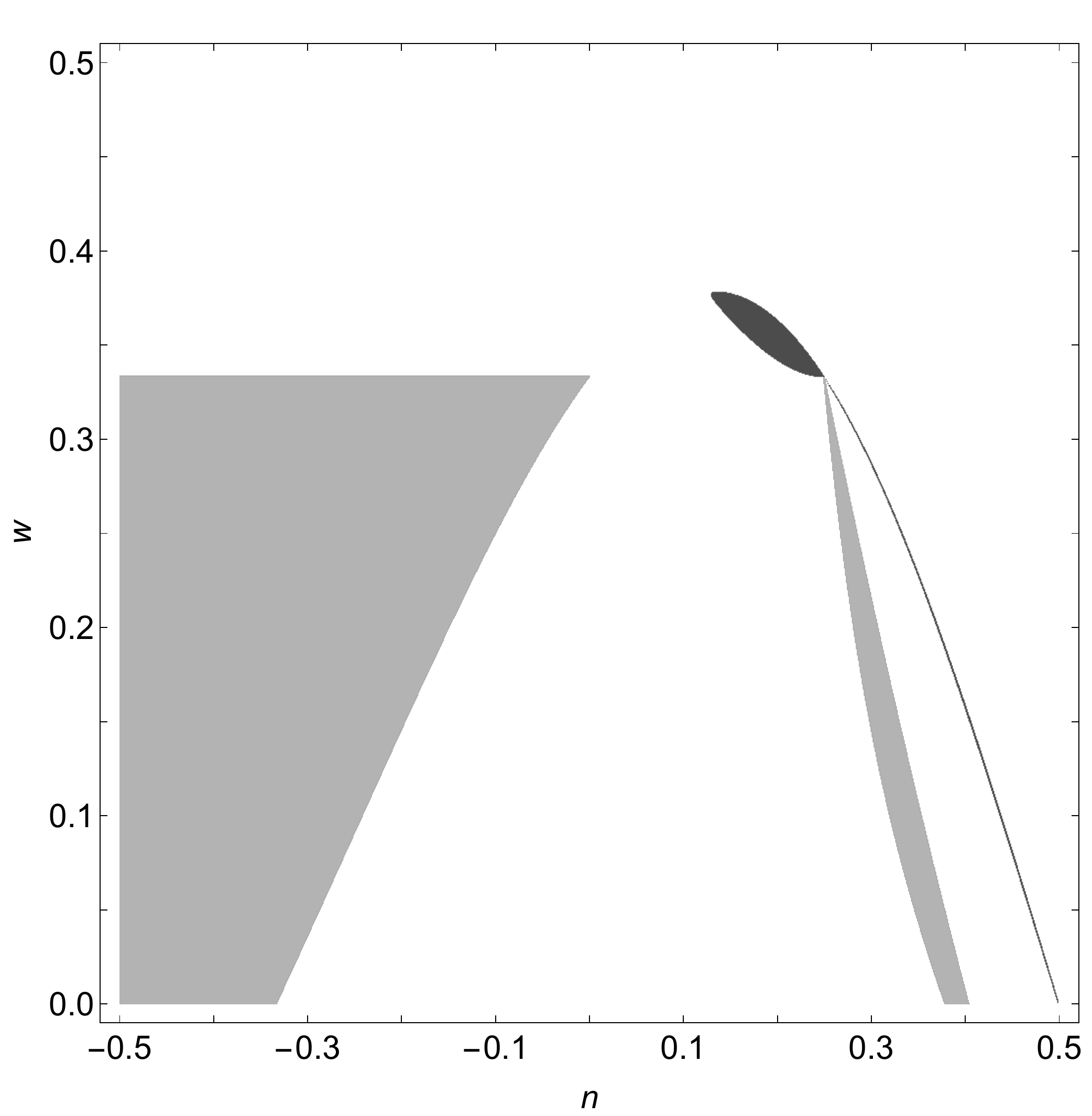}
	\caption[Stability regions of point $\ca$.]{Stability regions of point $\ca$. The dark grey area corresponds to an unstable region, while the light grey corresponds to a stable one. The remaining space corresponds to a saddle point. Note that the thin line to the right of the light grey area is a dark grey region with measurable width.\label{fig:powerfRA}}
\end{figure}

Point $\mathcal{A}$ is a point whose deceleration parameter depends on the type of matter present in the universe in the same way as in GR. It also requires that the scalar curvature and energy density be related by
\beq\label{rhor}
\rho=\kappa R	\fr{ n \left[6 n (w+1)+3 w+5\right]-2}{n \left[6 n (w+1)-3 w-1\right]+3 w-1}.
\eeq
Its stability regions can be seen in Fig. \ref{fig:powerfRA}.

While this point has several stable regions, they all require $w<-1/3$ in order to have an accelerated expansion of the universe (again, as in GR), and are thus unsuitable as a candidate for dark energy.

\subsubsection*{Point $\mathcal{B}$}\label{subsubsec:powerfRB}
\begin{figure}
	\centering
	\includegraphics[width=\columnwidth]{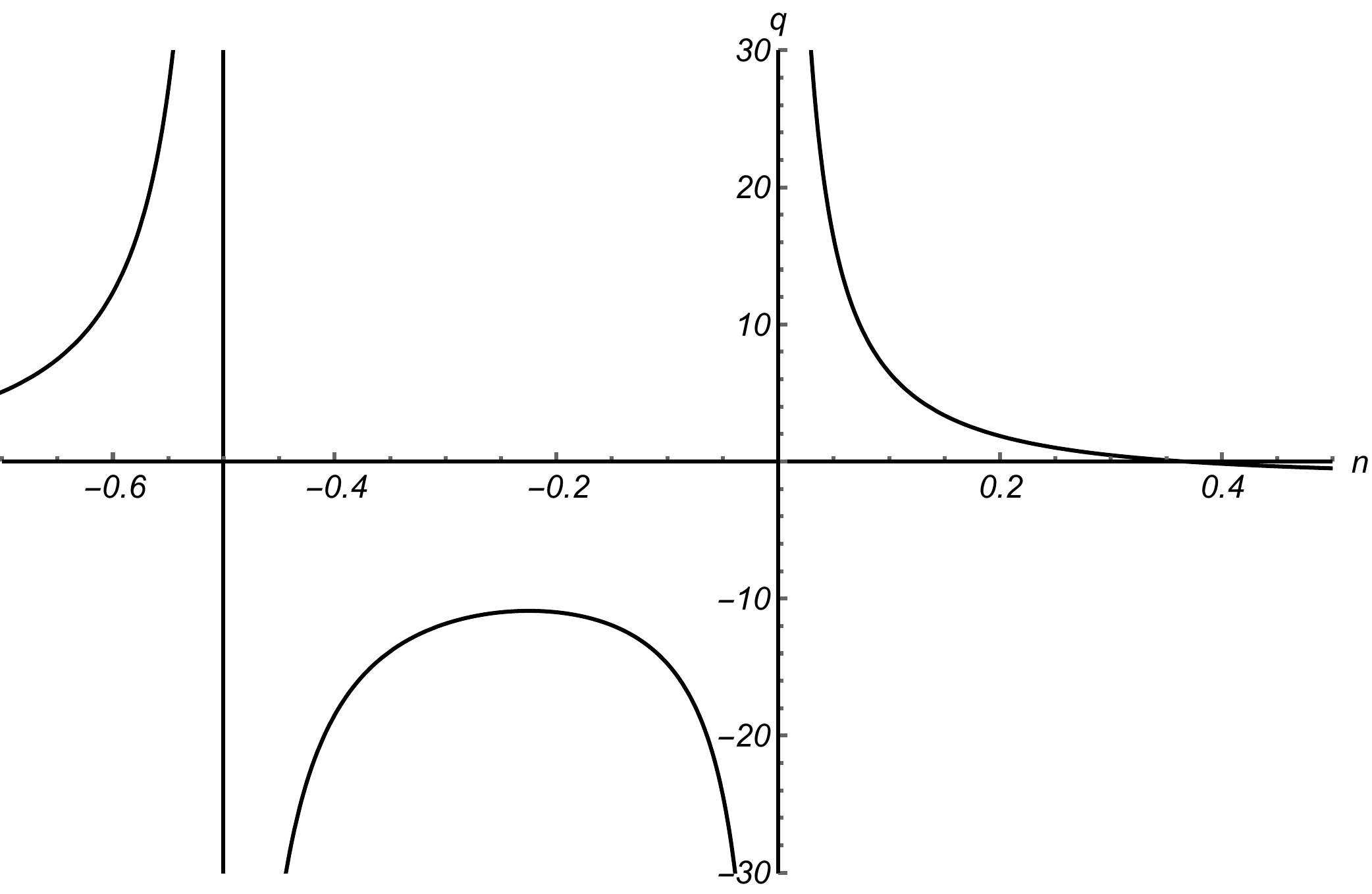}
	\caption[Deceleration parameter for point $\cb$.]{Deceleration parameter for point $\cb$ as a function of the exponent $n$.\label{fig:powerfRBq}}
\end{figure}

\begin{figure}
	\centering
	\includegraphics[width=\columnwidth]{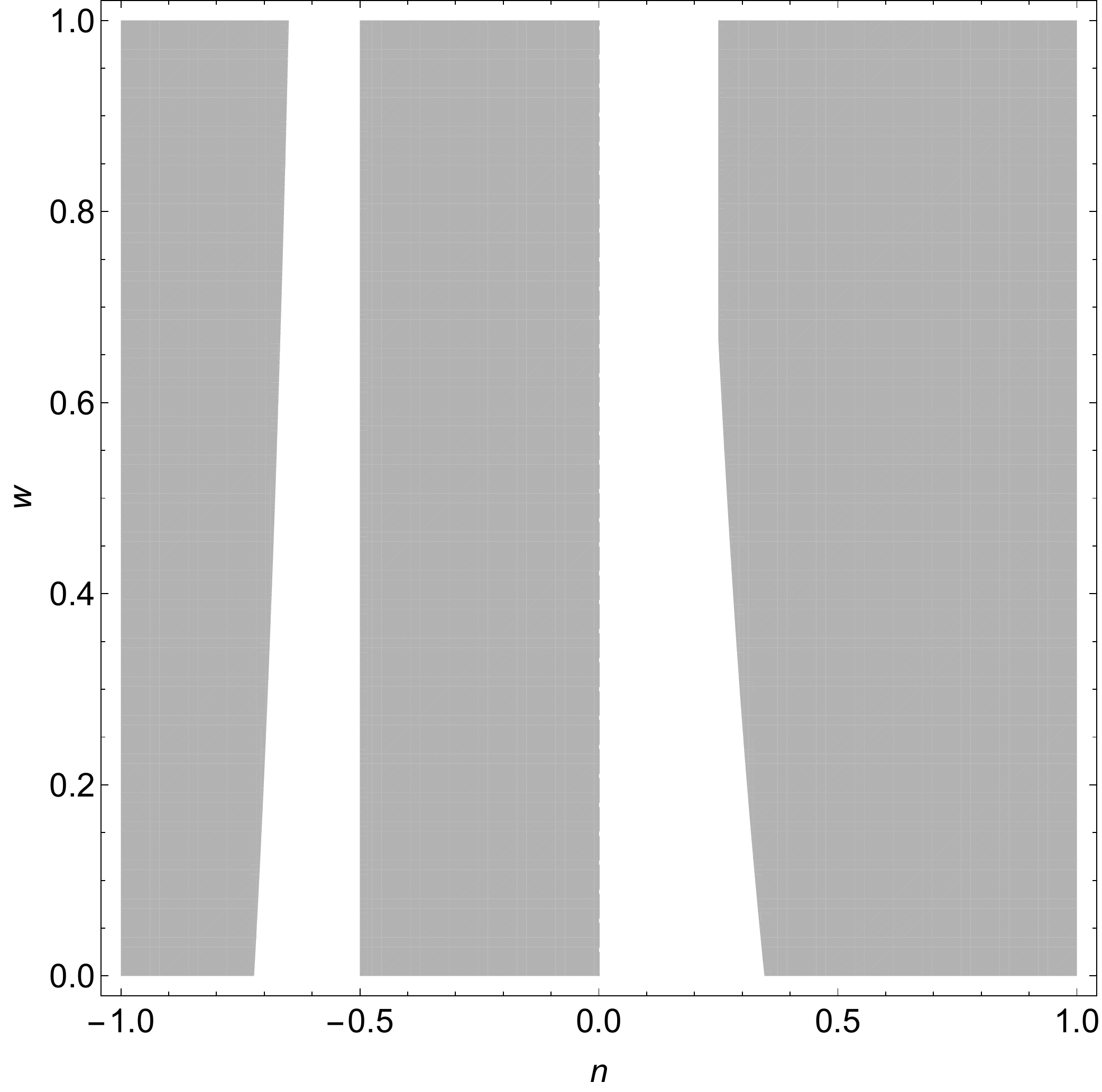}
	\caption[Stability regions of point $\cb$.]{Stability regions of point $\cb$. The light grey corresponds to a stable region and the remaining space corresponds to a saddle point.\label{fig:powerfRB}}
\end{figure}

This point corresponds to a universe with a vanishing energy density $\rho\rightarrow0$ and a deceleration parameter given by
\beq\label{decelB}
q=-1+\fr{1}{n}-\fr{3}{2 n+1},
\eeq
and depicted in Fig. \ref{fig:powerfRBq}.
The stability of this point can be found in Fig. \ref{fig:powerfRB}.
This point presents a viable candidate for dark energy, as it can be arbitrarily close to GR while still maintaining a negative deceleration parameter. In particular, if a linear coupling $f(R) \sim R $ is considered, the condition $f'_0 R_0 = f_0$ found in Subsection \ref{nonemptydeSitter} is fulfilled and a de Sitter phase is attained, $q = -1$.

Furthermore, this fixed point also includes the possibility of a ``big rip'' scenario, as $|q|$ can be arbitrarily large for the stable region $n \ll 1$.

Finally, it should be noted that, although this fixed point corresponds to a universe devoid of matter, $y=z=\rho =0$, it can nevertheless mimic the evolution of a matter-dominated universe as found in GR, {\it i.e.},
\beq
q = {1 +3 w \over 2} \to n = -{5+3w \pm \sqrt{73+78w + 9w^2} \over 12(1+w)}.
\eeq

\subsubsection*{Point $\mathcal{C}$}\label{subsubsec:powerfRC}
This solution is a saddle point with vanishing scalar curvature $R=0$, which also requires that $n$ and $w$ be related by $n=2/(4+3w)$. As $R=0$ and $y=2$ imply that $H=0$, this leads to a static universe and undefined deceleration parameter.

\subsubsection*{Point $\mathcal{D}$}\label{subsubsec:powerfRD}
Similarly to the previous point, point $\mathcal{D}$  has $R=0$ and $y\neq0$, implying a static universe and undefined deceleration parameter. Its stability can be found in Fig. \ref{fig:powerfRB}. It is interesting that this point presents a stable region for an as of yet unobserved evolution of the Universe, which could at first glance suggest that the current accelerated expansion phase is not the final stage in our Universe's evolution.
\begin{figure}
	\centering
	\includegraphics[width=\columnwidth]{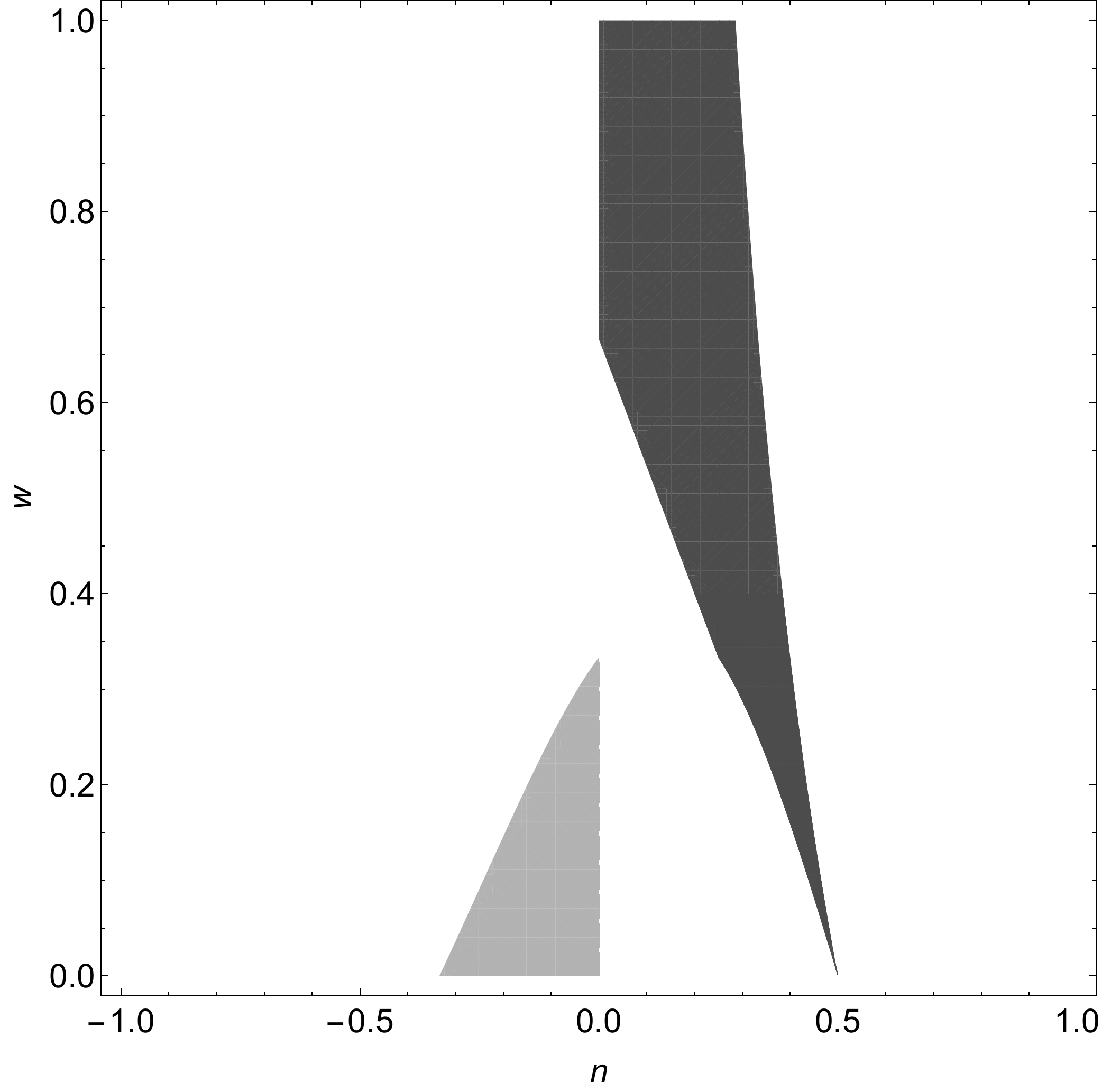}
	\caption[Stability regions of point $\cd$.]{Stability regions of point $\cd$. The dark grey area corresponds to an unstable region, while the light grey corresponds to a stable one. The remaining space corresponds to a saddle point.\label{fig:powerfRD}}
\end{figure}

\section{Discussion and outlook}\label{sec:conclusion}

In this work, the novel case of a model with Lagrangian density $f(R)\left(\kappa R+\cl\right)$ was presented, and an initial dynamical analysis was performed. While technically a particular case of NMC theories, the particular coupling presents several interesting solutions.

General conditions for the function $f(R)$ where obtained in order to allow for an accelerated expansion of the Universe: while the weaker relation $f'_0 R_0 = f_0$ is sufficient if matter is absent, a non-vanishing energy density requires the stronger conditions $f'_0=  f_0=0$. Notwithstanding the practical difficulty of realising a model fulfilling the latter, the possibility of having a Universe which might have a substantial matter content undergoing a de Sitter phase is alluring.

In order to further characterise the cosmology of the model under scrutiny, a dynamical system approach was first formulated and then applied to two natural candidates for the function $f(R)$: the ensuing results show that both exponential and power-law forms for the latter exhibit fixed points able to account for the current accelerated expansion of the Universe, as well as for inflation.

Even though such a dynamical system analysis proves itself to be extremely useful in cosmology, one must beware of several caveats inherent to its formulation: firstly, the system, and therefore its solutions, is dependent on the choice of variables, so one could omit interesting regimes purely by choosing a specific set of variables in favour of another. Secondly, the existence of any two fixed points for a given theory does not imply that they are connected by any type of trajectory, as noted in Ref. \cite{Carloni2007br}, and one cannot straightforwardly assume that any desirable attractor solution is in fact a global attractor, {\it i.e.} all trajectories will drive the universe towards that solution; as such, one may still be subject to a fine-tuning problem, which can only be ascertained with a further topological characterisation of the phase space of the model or an independent, direct integration of the equations of motion.
\appendix*
\section{Physical quantities}\label{app:physicalq}
Here are listed a few relevant physical quantities in terms of the used dimensionless variables \eqref{variables}. With the adopted metric \eqref{line}, the Ricci scalar reads
\begin{equation}\label{Ricci}
	R=6\left(2H^2+\dot{H}\right).
\end{equation}
\noindent One important parameter used in cosmology is the deceleration parameter
\begin{equation}\label{decel}
	q \equiv -{\ddot{a}a \over \dot{a}^2}=1-y,
\end{equation}
so that the scalar curvature may be written as
\begin{equation}
	R=6H^2(1-q).
\end{equation}
After determining the fixed points of the dynamical system for each particular choice of the function $f(R,\cl)$, we may straightforwardly determine the scale factor for each fixed point. From a direct integration of Eq. \eqref{decel} (for a fixed $y$), one obtains the general solution
\begin{equation}\label{scale}
	a(t)=
	\begin{cases}
		\left({t \over t_0}\right)^{1 \over 2-y}, & y \ne 2 \\
		e^{H_0 t}, & y=2
	\end{cases}.
\end{equation}
\noindent For the first case, the scale factor evolves as a power of time, while in the second result the Hubble parameter will be constant and this the scale factor will rise exponentially, i.e a De Sitter phase. Note that this solution was obtained resorting (indirectly) to the definition of the Ricci scalar with the used metric.

Other important physical quantity is the energy density: one can determine its evolution for each fixed point from the continuity Eq. \eqref{contin}. The general solution for this is the familiar result
\begin{equation}\label{density}
	\rho(t)=\rho_0 a(t)^{-3(1+w)}.
\end{equation}


\bibliographystyle{apsrev4-1}
\bibliography{newfR}

\end{document}